\documentclass{aa}  

\usepackage{graphicx}
\usepackage{txfonts}
\defcitealias{2015A&A...subm....H}{Paper~I}
\bibpunct[; ]{(}{)}{;}{}{}{,}  

\begin{document}
 
 \title{MUSE crowded field 3D spectroscopy of over 12\,000 stars in the globular cluster \object{NGC~6397}\thanks{Based on observations obtained at the Very Large Telescope (VLT) of the European Southern Observatory, Paranal, Chile (ESO Programme ID 60.A-9100(C))}}

 \subtitle{II. Probing the internal dynamics and the presence of a central black hole.}
 
 \author{S. Kamann\inst{1}
         \and
         T.-O. Husser\inst{1}
         \and
         J. Brinchmann\inst{2,3}
         \and
         E. Emsellem\inst{4}
         \and
         P.~M. Weilbacher\inst{5}
         \and
         L. Wisotzki\inst{5}
         \and
         M. Wendt\inst{5,6}
         \and
         D. Krajnovi\'c\inst{5}
         \and
         M.~M. Roth\inst{5}
         \and
         R. Bacon\inst{7}
         \and
         S. Dreizler\inst{1}
         }
         
  \institute{$^1$ Institut f\"ur Astrophysik, Georg-August-Universit\"at G\"ottingen, Friedrich-Hund-Platz 1, 37077 G\"ottingen, Germany\\
             $^2$ Leiden Observatory, Leiden University, PO Box 9513, 2300 RA Leiden, The Netherlands\\
             $^3$ Instituto de Astrof{\'i}sica e Ci{\^e}ncias do Espaço, Universidade do Porto, CAUP, Rua das Estrelas, PT4150-762 Porto, Portugal\\
             $^4$ ESO, European Southern Observatory, Karl-Schwarzschild Str. 2, 85748 Garching bei Muenchen, Germany\\
             $^5$ Leibniz-Institut für Astrophysik Potsdam (AIP), An der Sternwarte 16, D-14482 Potsdam, Germany\\
             $^6$ Institut f\"ur Physik und Astronomie, Universit\"at Potsdam, 14476 Potsdam, Germany\\
             $^7$ CRAL, Observatoire de Lyon, CNRS, Université Lyon 1, 9 avenue Ch. André, 69561 Saint Genis-Laval Cedex, France\\
}
 
 \date{Received; accepted}
 
 \abstract{We present a detailed analysis of the kinematics of the Galactic globular cluster \object{NGC~6397} based on more than ${\sim}18\,000$ spectra obtained with the novel integral field spectrograph MUSE. While \object{NGC~6397} is often considered a core collapse cluster, our analysis suggests a flattening of the surface brightness profile at the smallest radii. Although it is among the nearest globular clusters, the low velocity dispersion of \object{NGC~6397} of $<5\,\mathrm{km\,s^{-1}}$ imposes heavy demands on the quality of the kinematical data. We show that despite its limited spectral resolution, MUSE reaches an accuracy of $1\,\mathrm{km\,s^{-1}}$ in the analysis of stellar spectra. We find slight evidence for a rotational component in the cluster and the velocity dispersion profile that we obtain shows a mild central cusp. To investigate the nature of this feature, we calculate spherical Jeans models and  compare these models to our kinematical data. This comparison shows that if a constant mass-to-light ratio is assumed, the addition of an intermediate-mass black hole with a mass of $600\,M_\odot$ brings the model predictions into agreement with our data, and therefore could be at the origin of the velocity dispersion profile. We further investigate cases with varying mass-to-light ratios and find that a compact dark stellar component can also explain our observations. However, such a component would closely resemble the black hole from the constant mass-to-light ratio models as this component must be confined to the central ${\sim}5\arcsec$ of the cluster and must have a similar mass. Independent constraints on the distribution of stellar remnants in the cluster or kinematic measurements at the highest possible spatial resolution should be able to distinguish the two alternatives.}
 
 
 \maketitle
 
 \section{Introduction}
 \label{sec:intro}
 The internal kinematics of globular clusters have been the subject of extensive research for several decades. This seems remarkable, however, because from a theoretical point of view globular clusters have always been considered very simple systems: dynamically old, without any substantial amount of gas, star formation, or dark matter \citep[e.g.][also see Chapter 6.1 of \citealt{1998gaas.book.....B} for an overview of general properties of globular clusters]{1996ApJ...461L..13M}. Indeed, the processing rates of modern computers make it possible to follow the dynamical evolution of globular clusters with N-body simulations that contain a realistic number of particles \citep[e.g.][]{1999PASP..111.1333A}. Consequently, such models can make precise predictions, for example about the evolution of a cluster in the tidal field of the Milky Way \citep{2003MNRAS.340..227B} or the impact of different formation scenarios on the observable dynamics \citep{2015MNRAS.450.1164H}. However, a limiting factor in our understanding of the cluster dynamics has always been the lack of kinematical data to provide a comparison. The pioneering studies of \citet{1977AJ.....82..810D} or \citet{1979AJ.....84..752G} measured velocities for $10$--$100$ stars, and even in recent years spectroscopic samples containing $>1\,000$ stars per cluster have been the exception rather than the rule (but see  \citealt{2000AJ....119.1268G} or \citealt{2009A&A...493..947S}). Only recently, with the advent of high quality proper motions \citep[e.g.][]{2014ApJ...797..115B} or the commissioning of powerful integral field spectrographs, such as MUSE \citep{2014Msngr.157...13B}, this picture has started to change.
 
 Still some questions about the nature of globular clusters remain unanswered today. One of them focusses on the existence of intermediate-mass black holes (IMBHs) in their centres. With masses around $10^2$--$10^5$ solar masses, these black holes would lie in the desert between the two dominant populations of stellar-mass black holes on the one hand and supermassive black holes (SMBHs) with masses $>10^5$ solar masses on the other. The centres of globular clusters seem favourable places for the existence of IMBHs given that the clusters appear as scaled down versions of galactic bulges, which in turn display well-established scaling relations with the black holes they harbour in their centres \citep[e.g.][]{2000ApJ...539L..13G,2003ApJ...589L..21M,2013ARA&A..51..511K}. Different formation scenarios for SMBHs have been proposed \citep[see][]{1978PhyS...17..193R,2009MNRAS.396..343R}, some of which include the formation of massive seeds in dense star clusters, for example via runaway merging of cluster stars \citep{2002ApJ...576..899P}. Therefore, constraining the presence of IMBHs in globular clusters  reveals important information about the mass assembly of the most massive black holes known.
 
 The presence of an IMBH would also alter the evolution of the host star cluster itself. For example, a massive black hole can slow down the core collapse of the cluster as gravitational encounters with this massive black hole scatter stars on wider orbits. The same effect may also suppress mass segregation within the stellar population \citep{2008ApJ...686..303G}. \citet{2007MNRAS.374..857T} studied the effect of gravitational interactions between the black hole and binary stars and found that clusters harbouring an IMBH can be characterised by a large core to half-mass radius ratio. On the other hand, \citet{2007MNRAS.379...93H} and \citet{2011ApJ...743...52N} found no significant difference in this respect between simulated clusters with and without a central black hole. Instead, \citet{2011ApJ...743...52N} argued that a moderate logarithmic slope in the central surface brightness profile of a cluster is a strong hint of the presence of an IMBH.
 
 Observational evidence for the existence of IMBHs in globular clusters is still scant. A handful of detections have been reported, for example by \citet{2008ApJ...676.1008N} or \citet{2013A&A...552A..49L}. However, in some cases these detections have been contested \citep{2010ApJ...710.1063V,2013ApJ...769..107L} and to date no detection has been confirmed by the analysis of an independent data set. This lack of confirmation is crucial because of the different methods that are in use to search for IMBHs. All reported detections so far come from the spectroscopic analysis of the integrated cluster light. A common criticism about this approach is that the integrated light in a cluster is dominated by the brightest giant stars, hence the line broadening observed in the spectra is not representative of the whole stellar population. Alternatively, one can measure the velocity of individual stars, either spectroscopically \citep[e.g.][]{2002AJ....124.3255V} or using proper motions \citep[e.g.][]{2010ApJ...710.1032A}. These observations are extremely challenging because of the high stellar densities and rather low velocities involved. Consequently, many studies make use of the \textit{Hubble} space telescope (HST) or adaptive optics (AO). For example, \citet{2013ApJ...769..107L} used AO-assisted integral field spectroscopy in the near-infrared and measured velocities in the centre of \object{NGC~6388} via aperture spectroscopy around the brighter stars. \citet{2015arXiv150702813L} claim that this approach is biased because blends of unresolved or fainter neighbouring stars pull the measured velocities to the cluster mean, which explains the discrepancy to their measurements that suggest the presence of a black hole.

 In \citet{2013A&A...549A..71K}, we introduced the approach of crowded field 3D spectroscopy to perform single star spectroscopy in integral field data of crowded stellar fields. This technique uses the point spread function of the observations to deblend the overlapping images of adjacent stars and to optimally extract the spectrum of each star. Hence, the extracted signal should be very robust against contamination from nearby sources. The potential of this approach was illustrated in \citet{2014A&A...566A..58K}, where we obtained stringent upper limits for an IMBH in three clusters using the PMAS spectrograph \citep{2005PASP..117..620R}, although the mediocre seeing and  small field of view only allowed us to observe a limited number of ${\sim}100$ stars per cluster.
 
In \citet[][hereafter \citetalias{2015A&A...subm....H}]{2015A&A...subm....H}, we presented crowded field 3D spectroscopy for a large number of stars in the globular cluster \object{NGC~6397} based on MUSE commissioning data. Using our source deblending algorithm, we extracted more than 18\,000 spectra, which were analysed with full spectral fitting against synthetic PHOENIX spectra from the library of \citet{2013A&A...553A...6H}. We use the radial velocity measurements obtained in the course of this analysis to investigate the internal dynamics of the cluster.
 
 \object{NGC~6397} has a moderate stellar mass. \citet{2012ApJ...761...51H} estimate that it measures $1.1\times10^5M_\odot$, which  results in a short relaxation time in combination with the
high stellar density. The database of \citet[][2010 edition]{1996AJ....112.1487H} gives a median relaxation time of $4\times10^8\,\text{years}$, which is much less than any age estimate of Galactic globular clusters. Consequently, strong signs of mass segregation have been found \citep[e.g.][]{1995ApJ...452L..33K,2014MNRAS.442.3105M}. A late evolutionary stage may also be mirrored by the surface brightness profile of \object{NGC~6397}. Its steep central slope is often considered as evidence that the cluster has undergone core collapse \citep[e.g.][]{1986ApJ...305L..61D}. In view of its photometric properties (possible core collapse, small core radius), \object{NGC~6397} does not appear to be a prime candidate to search for an IMBH. On the other hand, for an unbiased view of the whole Galactic globular cluster population in this respect, it is important to also obtain constraints for such ``unfavourable'' clusters.
 The paper is structured as follows. Following a brief presentation of the MUSE data in Sect.~\ref{sec:dataset}, we analyse the surface brightness profile of \object{NGC~6397} in Sect.~\ref{sec:sbprofile}. Afterwards, a careful judgement of the quality of the radial velocity measurements derived from the MUSE data is performed in Sect.~\ref{sec:radvel}. A cleaning of our sample from non-member stars is performed in Sect.~\ref{sec:membership} before the analysis of the velocity dispersion profile is presented in Sect.~\ref{sec:kinematics}. We conclude in Sect.~\ref{sec:conclusions}. Throughout this work, we provide magnitudes in the Vega system. Radial velocities are provided in a heliocentric system.

 \section{Data set}
 \label{sec:dataset}
 The data that this study is based on was obtained in August 2014 during the commissioning of MUSE \citep{2014Msngr.157...13B}, the new panoramic integral field spectrograph at the ESO/VLT. The observations covered the central part of \object{NGC~6397} with a mosaic of $23$ pointings, each one covering an area of $1\arcmin\times1\arcmin$ on the sky. For seven pointings repeated observations were performed during a following night. The exposure times were short, typically one minute, to avoid saturating the brightest stars. However, each pointing was observed in a 4-point dither pattern with derotator offsets of $90$ degrees to average out possible systematic effects from individual spectrographs. The wavelength range of MUSE in the nominal mode used during the observations was $4800$--$9300\,{\text \AA}$ with a spectral resolution of $R{\sim}1\,700$ in the blue and $R{\sim}3\,500$ in the red.
 
 The data reduction, the extraction of the individual stellar spectra, and their analysis are described in \citetalias{2015A&A...subm....H}, which provides details of these steps. Here we restrict ourselves to a summary of the main characteristics of the extracted spectra. The final sample consists of $18\,932$ spectra for $12\,307$ stars. For $5\,613$ stars, more than one spectrum was extracted, either because the stars were situated in the overlap region of adjacent pointings or because the pointing was observed during two nights. The faintest stars in the sample have a magnitude of ${\rm V}{\sim}19.5$, which are approximately $5~{\rm mag}$ fainter than the main sequence turn-off of \object{NGC~6397}. The vast majority ($87\%$) of our sample are main-sequence stars, and the remaining objects are either giants or horizontal branch stars. The signal-to-noise (S/N) of the extracted spectra varies with the brightnesses of the stars. Spectra of giant stars were typically extracted with ${\rm S/N}>100$. Going to fainter magnitudes, the average S/N reduces to ${\sim}50$ around the main sequence turn-off, degrades further to ${\sim}20$ at a magnitude of ${\rm V}{\sim}18$, and approaches unity for the faintest stars in our sample. In total, $10\,521$ spectra have ${\rm S/N}>10$.
 
 \section{Surface brightness profile}
 \label{sec:sbprofile}
 
 \begin{figure}
  \includegraphics[width=\hsize]{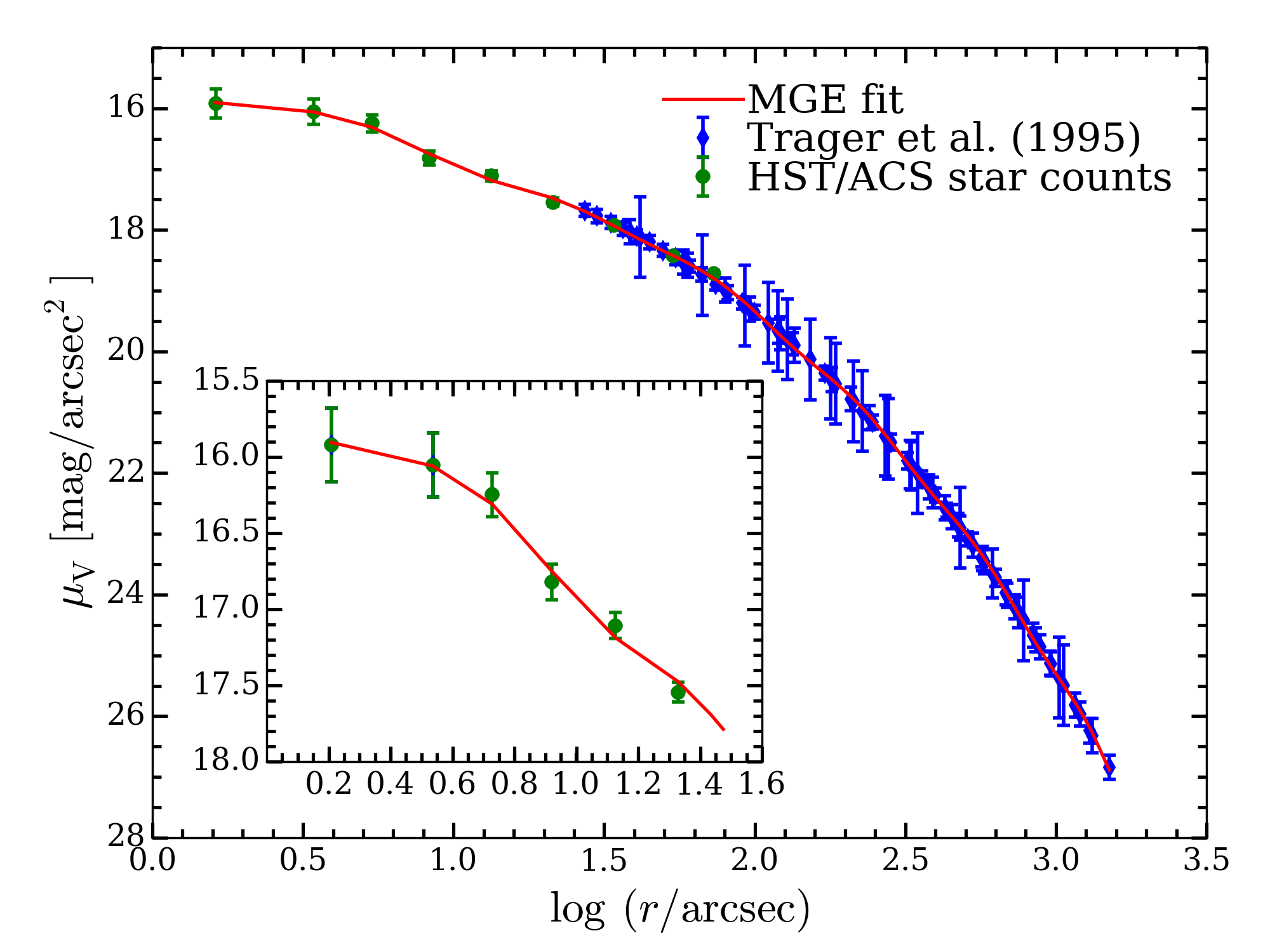}
  \caption{Surface brightness profile of \object{NGC~6397} across the entire extent of the cluster. The insert shows a zoom into the central region. Blue diamonds show the data of \citet{1995AJ....109..218T}, green circles were obtained from star counts using an HST/ACS catalogue, and the red line shows a fit using the MGE method with six Gaussian components.}
  \label{fig:sbprofile}
 \end{figure}
 
 For any dynamical modelling of a globular cluster, an accurate measurement of the surface brightness profile is crucial. Its deprojection yields the three-dimensional stellar density, which is  key to inferring the gravitational potential of the (bright) stars and to obtaining observable kinematic tracers of stars moving in this potential.
 
 To obtain the surface brightness profile of \object{NGC~6397}, we followed the same approach as in \citet{2014A&A...566A..58K}. Star counts from HST/ACS F606W data were used to obtain the central profile. The ACS data, obtained in the Wide Field Channel (WFC), only reach out to ${\sim}2\arcmin$ from the cluster centre, and  our MUSE data cover just a slightly larger area. Therefore, we complemented the star counts with the archival data of \citet{1995AJ....109..218T}, which represent a compilation of ground-based measurements from various telescopes and cover the entire extent of the cluster ($r\lesssim30\arcmin$). In addition, the \citeauthor{1995AJ....109..218T} profile served as calibration data for the star counts in the sense that the profile obtained from the star counts was shifted vertically until it matched the \citeauthor{1995AJ....109..218T} profile in the radial range where the two profiles overlapped. 
 An alternative to star counts would have been the usage of the high-resolution profile obtained by \citet{2006AJ....132..447N} to model the inner profile. However, the advantage of the star counts approach is that we can obtain the profile for approximately the same stellar subsample for which we have kinematical data. The faintest stars for which we obtained useful radial velocities with MUSE are at $\text{V}{\sim}19$ and the data are fairly complete down to $\text{V}=18$, hence we used the latter value as our magnitude cut. The input catalogue was obtained as part of the ACS survey of Galactic globular clusters \citep{2007AJ....133.1658S,2008AJ....135.2055A} and is the same catalogue we used in \citetalias{2015A&A...subm....H} to extract the stellar spectra. Artificial star tests are provided as part of the survey data and we used them to check the completeness of the catalogue. At $\text{V}=18$, the catalogue is $>95\%$ complete across the whole footprint of the observed field, therefore, we do not expect any incompleteness-related flattening of the profile towards the cluster centre.
 
 We counted stars in concentric annuli around the cluster centre, which we assumed to be at the location determined by \citet{2010AJ....140.1830G}. The size of each bin was chosen so that it contained at least $20$ stars above our brightness cut and covered at least a radial range of $d\log (r/\arcsec)=0.2$. The largest annulus had an outer radius of $80\arcsec$. From the \citet{1995AJ....109..218T} data, we used the measurements with distances $>25\arcsec$ to the cluster centre.
 
 For further analyses, the profile was fitted with the multi-Gaussian expansion (MGE) method \citep{1994A&A...285..723E} via the software package of \citet{2002MNRAS.333..400C}, which fits the surface brightness profile with a sequence of Gaussians of different widths. In our case, a total of six Gaussians provided a satisfactory fit to the profile. The MGE fit together with the observed data is shown in Fig.~\ref{fig:sbprofile}.
 
 We determined the apparent integrated magnitude of the cluster in two ways: once by summing up the contributions of all Gaussians in the MGE fit and another time through direct numerical integration of the measured profile. Both methods gave completely consistent results of $V_\mathrm{tot, data} = 6.02$ and $V_\mathrm{tot, fit} = 6.00$, respectively, again highlighting the agreement between the data and the fit. The formal uncertainties of these values, obtained via jackknife, are small, $0.01\,{\rm mag}$. However, literature estimates of the integrated magnitude of \object{NGC~6397} show a large variance, ranging from $V_\mathrm{tot}=5.17$ \citep{2012AJ....144..126D} to $V_\mathrm{tot}=6.20$ \citep{1991ApJ...375..594V}. This scatter may be explained by the sensitivity of the integrated magnitude on the shape of the adopted analytical profile. \citet{2012AJ....144..126D} found differences of up to $0.5\,{\rm mag}$ when comparing the results from \citet{1966AJ.....71...64K} and \citet{1975AJ.....80..175W} profiles. In principle, the MGE approach that we adopted should be robust against such errors because no specific shape of the profile is required. However, one uncertainty that remains is the absolute zero point of our surface brightness profile, which solely depends on the data of \citet{1995AJ....109..218T}. We discuss how this uncertainty affects our results in \ref{sec:jeans}.
 
 \citet{2007ApJ...671..380H} determined the true distance modulus of \object{NGC~6397} to $m-M=12.02$. Using this value and accounting for an extinction of $A_\text{V}=0.56$ \citep[assuming $A_\text{V} = 3.1\cdot E_\text{B-V}$ and using $E_\text{B-V}=0.18$,][]{2003A&A...408..529G}, we obtain an absolute magnitude of the cluster of $M_\mathrm{V}=-6.57$.
 
 The zoom into the central region of the surface brightness profile in Fig.~\ref{fig:sbprofile} shows the steep rise towards the centre, which is typical for globular clusters that have undergone core collapse. Still, a possible flattening within ${\sim}2\arcsec$ of the profile may be indicated by the innermost data point. In this respect, it is interesting that \citet{2006AJ....132..447N} speculate whether \object{NGC~6397} is actually a core-collapse cluster. These authors find that models with small cores of ${\sim}2\arcsec$--$5\arcsec$ provide good fits to the central part of the surface brightness profile, which qualitatively agrees with our finding of a potential flattening. The question of whether \object{NGC~6397} has undergone core collapse was also discussed by \citet{1991A&A...250..113M}, who argued that the low number of bright stars in the very central region makes any classification doubtful unless high quality profiles based on star count are obtained.
 
 The evolutionary state of \object{NGC~6397} also relates to the possible presence of an intermediate-mass black hole. Simulations carried out by \citet{2005ApJ...620..238B} or \citet{2011ApJ...743...52N} showed that core-collapse clusters are unlikely to contain massive black holes. Instead, the photometric fingerprint of an IMBH would be a shallow cusp $\mu(r)\propto r^{-\gamma}$, where a range of $0.1<\gamma<0.4$ was obtained by \citet{2011ApJ...743...52N}. The value of $\gamma=0.37$ obtained by \citet{2006AJ....132..447N} would thus still fall into the ``allowed'' range.
 
 \section{Radial velocities}
 \label{sec:radvel}
 
 In \citetalias{2015A&A...subm....H}, we analysed $18\,932$ spectra for $12\,307$ stars in a $5\arcmin\times5\arcmin$ field centred on NGC~6397. Radial velocities were measured during a full-spectrum fit of synthetic templates, from the library presented in \citet{2013A&A...553A...6H}, to the MUSE spectra. From this sample, we selected all measurements that were derived from spectra extracted with an average $\mathrm{S/N}>10$ across the full wavelength range. This selection left us with $10\,521$ radial velocities for $7\,131$ stars. As a result of the decrease in stellar density with distance to the cluster centre, we resolved a larger fraction of relatively faint stars in the outskirts of our sample than in the centre. As many of the extracted spectra of these faint stars have a low S/N, the fraction of excluded spectra was higher in the outskirts of the sample.

 Cross-correlation, either against synthetic or observed templates, would be an alternative approach to obtain radial velocities. As a consistency check, we correlated all extracted spectra against matched templates (based on $[\text{M/H}]$, $T_\text{eff}$, and $\log g$) from the same library. After correcting for a small global offset of ${\sim}2\,\mathrm{km\,s^{-1}}$, we achieved consistent results, i.e. the scatter was explainable by the measurements uncertainties. However, the uncertainties for the cross-correlation measurements were larger by ${\sim}30\%$. We speculate that the gain in accuracy in the spectral fit originates from the additional adaptions of the template spectra, such as  inclusion of telluric absorption or matching of the line widths. We use the results from the spectral fit in the following.
 
\begin{figure}
  \includegraphics[width=\hsize]{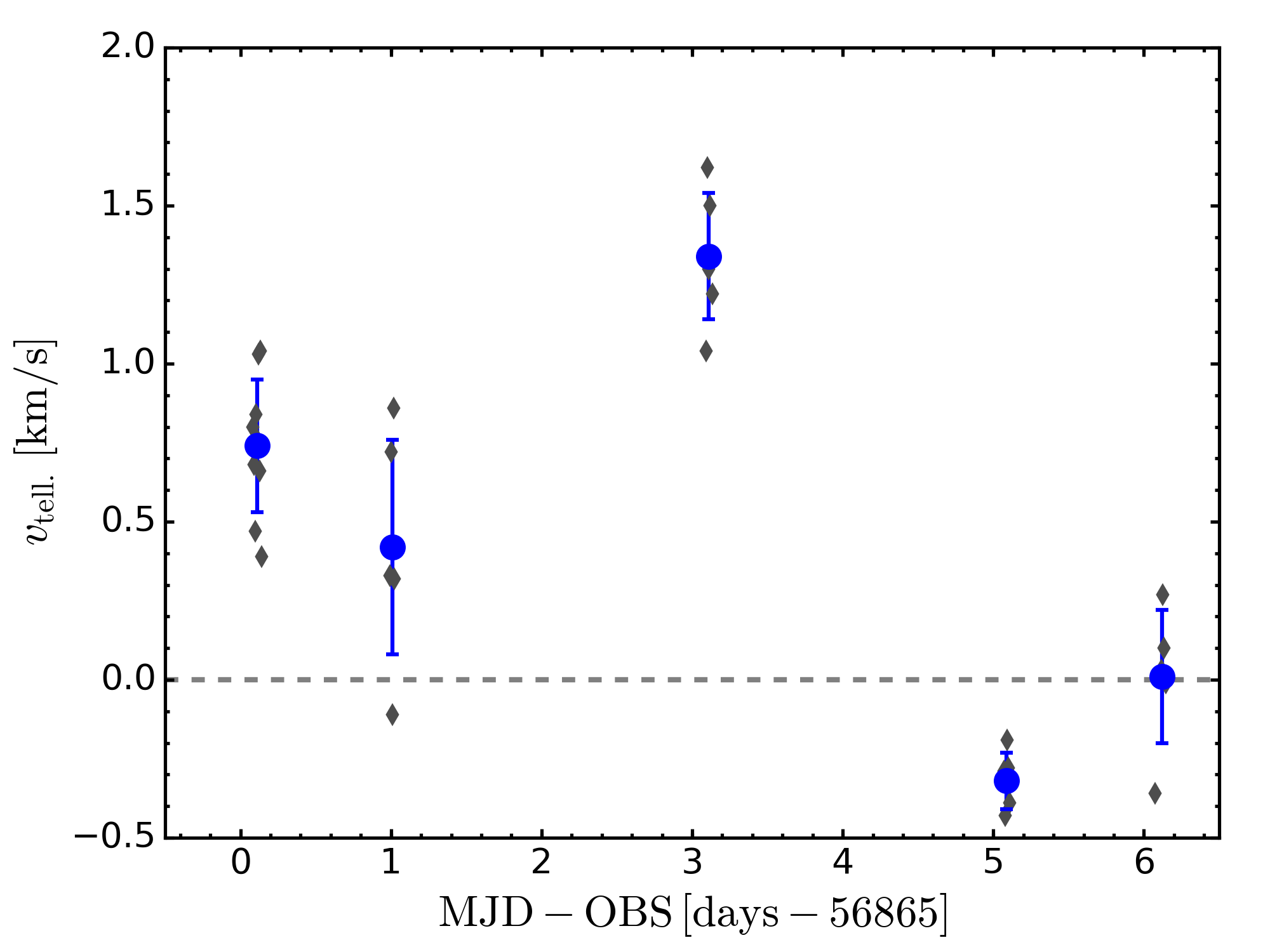}
  \caption{Mean velocity of the telluric component in the spectral fit as a function of observing night. Dark grey diamonds represent individual datacubes and were obtained by averaging the velocities measured for the telluric absorption bands of all spectra extracted with a S/N > 30 from this cube. The blue circles and associated error bars were obtained by averaging the results from all datacubes observed in a single night.}
  \label{fig:tellurics}
 \end{figure}

 In the low velocity regime that we face in NGC~6397, it is important that all measured radial velocities are in the same reference system. Our data were observed over the course of a week and slight changes in the wavelength calibration could already lead to a spurious rotational signal in the cluster or a systematic overestimation of the velocity dispersion. To check the accuracy of our wavelength calibration, we used the velocities measured for the telluric absorption bands during the spectral fit. In Fig.~\ref{fig:tellurics}, we show  the average velocity of the telluric component in all spectra with a S/N > 30 for all the datacubes that we analysed. Considering the complexity of MUSE and the rather low spectral resolution of the instrument ($R{\sim}2\,000\,$--\,$4\,000$), the accuracy that we achieve is very remarkable. The night-to-night variation in the velocity zeropoints is ${\sim}2\,\mathrm{km\,s^{-1}}$ at most, while during a single night we typically achieve an accuracy of $1\,\mathrm{km\,s^{-1}}$ or better.   We determined the average velocity for each night (blue circles in Fig.~\ref{fig:tellurics}) and subtracted this velocity from all measurements taken during that night to correct for the offsets measured during different nights. We considered the scatter observed in a single night as our final wavelength inaccuracy and added $1\,\mathrm{km\,s^{-1}}$ in square to all our velocity uncertainty measurements.
 
 \subsection{Uncertainties}
 \label{sec:uncertainties}
 
\begin{figure}
 \includegraphics[width=\hsize]{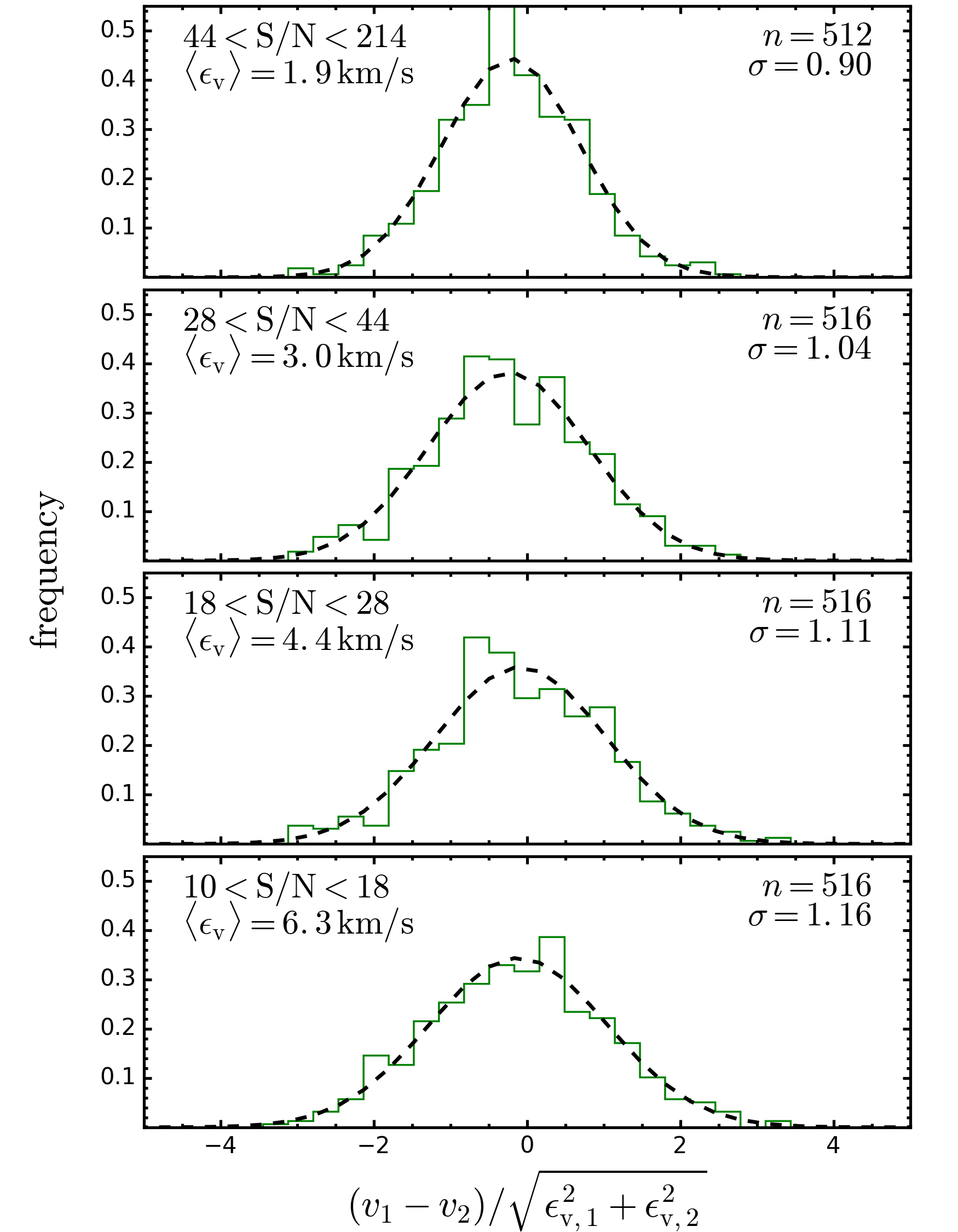}
 \caption{Measured radial velocity differences for stars with multiple measurements. Different histograms correspond to different S/N ratios per pixel of the analysed spectra, as indicated in the upper left of each panel together with the average uncertainty of the contributing velocity measurements. The black dashed lines show Gaussian fits to the histograms, the number of stellar pairs, and the standard deviation of each fitted Gaussian are given in the top right of each panel.}
 \label{fig:uncertainties}
\end{figure}

 A proper calibration of the measurement uncertainties is crucial for all future investigations. It is well known \citep[e.g.][]{2006A&A...445..513V} that improperly calibrated uncertainties can lead to wrong estimates of the velocity dispersion and hence to wrong model parameters. In this respect, the low velocity dispersion of \object{NGC~6397}, $\sigma_\mathrm{c}{\sim}4.5\,\mathrm{km\ s}^{-1}$ \citep{1996AJ....112.1487H}, is a challenge because the squared ratio of the uncertainties and the intrinsic dispersion determines the influence of the former.
 In our case, the uncertainties were derived from the covariance matrix returned by the Levenberg-Marquardt algorithm that implemented the spectral fit. To calibrate them, we make use of the stars that have been observed multiple times. Among these, we searched for pairs of spectra with comparable S/N ratios and compared the velocity difference $v_\text{2}-v_\text{1}$ to the uncertainties of the individual measurements, $\epsilon_{\text{v},1}$ and $\epsilon_{\text{v},2}$. In case of correct uncertainties, the quantity
 \begin{equation}
  \delta v_\text{1,2} = \frac{v_\text{2}-v_\text{1}}{\sqrt{\epsilon_{\text{v},1}^2 + \epsilon_{\text{v},2}^2}}
 \end{equation}
should be normally distributed with a standard deviation of one\footnote{As already mentioned in \citetalias{2015A&A...subm....H}, the assumption of a Gaussian distribution is only valid if the uncertainties themselves have negligible uncertainties. Otherwise, a Student-t distribution must be used. However, we find that for our data Gaussians provide close representations of the measured distributions (cf. Fig.~\ref{fig:uncertainties})}. A narrower distribution would hint at uncertainties that are systematically too high, while underestimated uncertainties would result in a wider distribution. In Fig.~\ref{fig:uncertainties}, histograms of $\delta v_\text{1,2}$ are shown for four S/N bins. In each bin, we considered measurement pairs stemming from spectra with S/N ratios that agreed within $25\%$ and have a mean value inside the indicated range. A single Gaussian was fitted to the data in each bin after removing obvious outliers, that is probably radial velocity (RV) variable stars, via $3\sigma$ clipping. 
 
 In general, the measured uncertainties provide an accurate description of the true uncertainties, however, there is a slight trend that the measured uncertainties are too small for stars with a low S/N and too high for the highest S/N. To account for this, we applied a correction factor to the uncertainties that was scaled with the S/N such that the distributions had a standard deviation of one after the  correction.
 
 A possible drawback of this approach is that RV variable stars systematically widen the distributions. However, the fraction of binary stars in \object{NGC~6397} has been found to be low (see Sect.~\ref{sec:rvvariables}) and the time spans between observations of the same field were short, typically ${\sim}24\mathrm{h}$. Therefore, we do not expect RV variables to have a significant impact on the widths of the distributions.
 
 \subsection{Literature data}
 \label{sec:literature}
 Our observations completely cover the central part of NGC~6397, out to ${\sim}2.5\arcmin$ distance from the cluster centre.  We complemented the MUSE data with literature data to cover the outer regions of the cluster as well. To this aim, we used the data sets of \citet{2009A&A...503..545L} and \citet{2009A&A...505..117C}.  Uncertainties are not given for either of the studies, however, since the radial velocities were derived from high-resolution spectra, we consider them to be accurate. To obtain average uncertainties, we used the stars that are also available in the MUSE data (based on their RA- and Dec-coordinates and $V$-band magnitudes) and looked at the scatter between our velocity measurements and those in the literature. To explain the scatter, average uncertainties of $0.8\,\mathrm{km\,s^{-1}}$ for the \citeauthor{2009A&A...505..117C} sample and $1.3\,\mathrm{km\,s^{-1}}$ for the \citeauthor{2009A&A...503..545L} sample are required. A third sample from \citet{1995AJ....110.1699G} is entirely covered by the MUSE observations, and the velocity measurements are consistent between both samples; only the Fabry-Perot measurements have higher uncertainties than our data, i.e. ${\sim}1.5\,{\rm km\,s^{-1}}$ for the brightest giants and ${\sim}4.0\,{\rm km\,s^{-1}}$ for the faintest stars in the sample with magnitudes $V{\sim} 16$. Still we used the Fabry-Perot data  to search for RV variable stars.

 \subsection{RV variable stars}
 \label{sec:rvvariables}
 When combining the individual measurements for each star, we flagged stars that showed RV variablility using the same criterion as in \citet{2014A&A...566A..58K}, namely a probability $<1\%$ that the scatter in the measurements is consistent with the uncertainties. This yielded 150 RV variable stars out of 2951 stars with multiple measurements. The low fraction of RV variable stars is in line with photometric searches for binary stars in this cluster, which typically yield fractions $<10\%$ \citep{2012A&A...540A..16M,2015ApJ...807...32J}. All stars flagged as RV variable were excluded from the analysis presented in Sect.~\ref{sec:kinematics}. However, we verified that the main results presented in that section did not depend on whether these stars were excluded or not.
 
\section{Membership}
\label{sec:membership}
 
 \begin{figure}
  \includegraphics[width=\hsize]{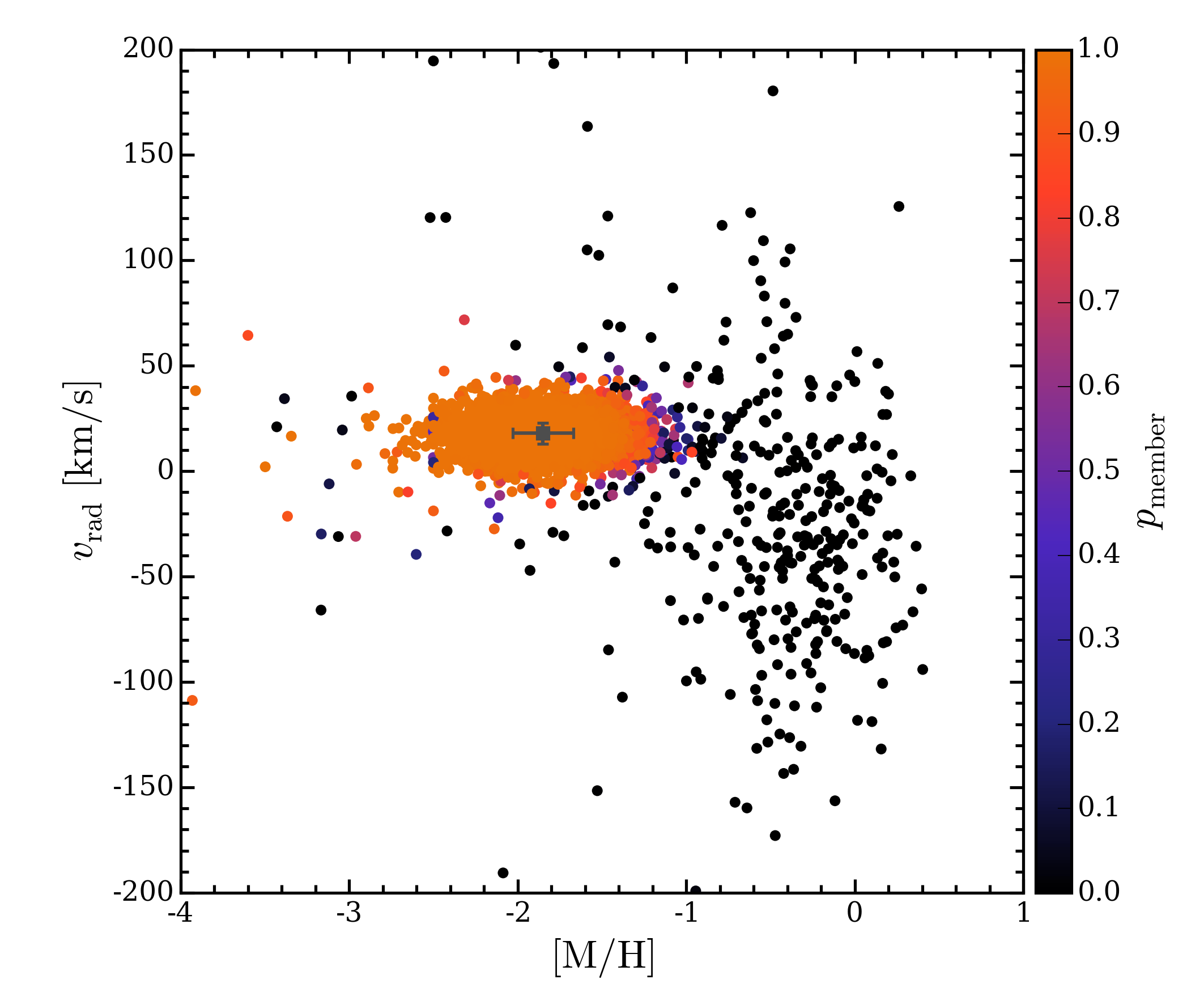}
  \caption{Distribution of all stars in the MUSE sample in the radial velocity vs. metallicity plane, colour-coded using their membership probabilities. Membership probabilities were determined using the EM technique as described in the text. For the cluster population, the mean radial velocity and metallicity plus their intrinsic dispersions are indicated with the grey square. The bulk of orange coloured member stars does not hide any non-member stars.}
  \label{fig:membership}
 \end{figure}
 
 As shown in \citetalias{2015A&A...subm....H}, our sample is affected by contamination of stars not associated with \object{NGC~6397}. This is to be expected in a blind spectroscopic survey across several arcmin on the sky. A sophisticated approach to identify a target population of stars against a background of Milky Way stars has been presented by \citet{2009AJ....137.3109W}, which we slightly modified according to the properties of our data. The \emph{expectation maximisation} (EM) technique iteratively determines the membership probability of each star in the data sample based on assumed distribution models of the target and the contaminating populations. 
 
 Our aim is to discriminate the two populations based on our measurements of the radial velocity $v_\text{r}$ and the metallicity $[\text{M/H}]$. As a model for the contaminating population, we adopted the Besancon numerical model of the Milky Way \citep{2003A&A...409..523R}. The quantities obtained from this simulation are denoted by a subscript \emph{bes} in the following. This simulation provides predictions of $v_\text{r, bes}$ and $[\text{M/H}]_\text{bes}$ for each star from a simulated Milky Way population along the line of sight towards the globular cluster. We modified the equation for the probability density of contaminating stars \citep[eq.~8 in][]{2009AJ....137.3109W} to be written as
 \begin{align}
  p_\text{non} \left(v_\text{r}, w=[\text{M/H}]\right)\nonumber &= \frac{1}{2\pi\cdot N_\text{bes}\cdot\epsilon_\text{v,bes}\cdot\epsilon_\text{w,bes}}\nonumber \\
  \times\sum_{i=1}^{N_\text{bes}}\exp&\left[-\frac{1}{2}\left(\frac{(v_{\text{bes},i}-v_\text{r})^2}{\epsilon_\text{v,bes}^2} + \frac{(w_{\text{bes},i}-w)^2}{\epsilon_\text{w,bes}^2}\right)\right]\,,
 \end{align}
 where $N_\text{bes}$ is the number of stars returned by the simulation. This implies that the probability density of observing a star from the contaminant population is calculated from a sum of $N_\text{bes}$ 2D Gaussian kernels. The kernels have constant widths, $\epsilon_\text{v,bes}$ and $\epsilon_\text{w,bes}$, for all stars. We set $\epsilon_\text{v,bes}=5\,\mathrm{km/s}$ and $\epsilon_\text{w,bes}=0.1$, matched to the uncertainties in the MUSE data.

 The probability density of the cluster population was modelled using a single 2D Gaussian distribution in $v_\text{r}$ and $[\text{M/H}]$ with the initial guesses for the mean values and widths of the distribution matched to literature values for \object{NGC~6397}.
 
 During the iteration, the membership probability for each measured star is determined under the prerequisite that the member fraction of the data decreases with increasing distance to the cluster centre. The task of finding the most likely function that is non-increasing can be tackled using monotonic regression. We used the isotonic regression model provided in the python \emph{sklearn} package \citep{scikit-learn} for this purpose.
 
 In Fig.~\ref{fig:membership} we show the membership probabilities of all stars from our MUSE sample in the $v_\text{r}$--$[\text{M/H}]$ plane. The EM technique seems to do a very good job of removing Milky Way stars from our cluster sample. As expected for a cluster at low Galactic latitude, the contaminant population mainly consists of metal-rich ($[\text{M/H}]>-1$) stars. The mean radial velocity for the cluster population is $17.8\,\mathrm{km\,s^{-1}}$ with a dispersion of $5.0\,\mathrm{km\,s^{-1}}$, while for the metallicity we measure a mean of $-1.89$ and a dispersion of $0.17$. These values are in good agreement with the literature and the results obtained in \citetalias{2015A&A...subm....H}.
 
 It is remarkable that only few stars fall in the boundary region between secure cluster members ($p=1$) and non-members ($p=0$), so that kinematic properties derived for the cluster in the following should not be sensitive to the adopted membership threshold. In the following, we  use all stars with $p\geq0.8$ to analyse the dynamics of the cluster.

 \section{Cluster kinematics}
 \label{sec:kinematics}
 
 \subsection{Velocity dispersion profile}
 \label{sec:dispersion}
 
 \begin{figure}
   \includegraphics[width=\hsize]{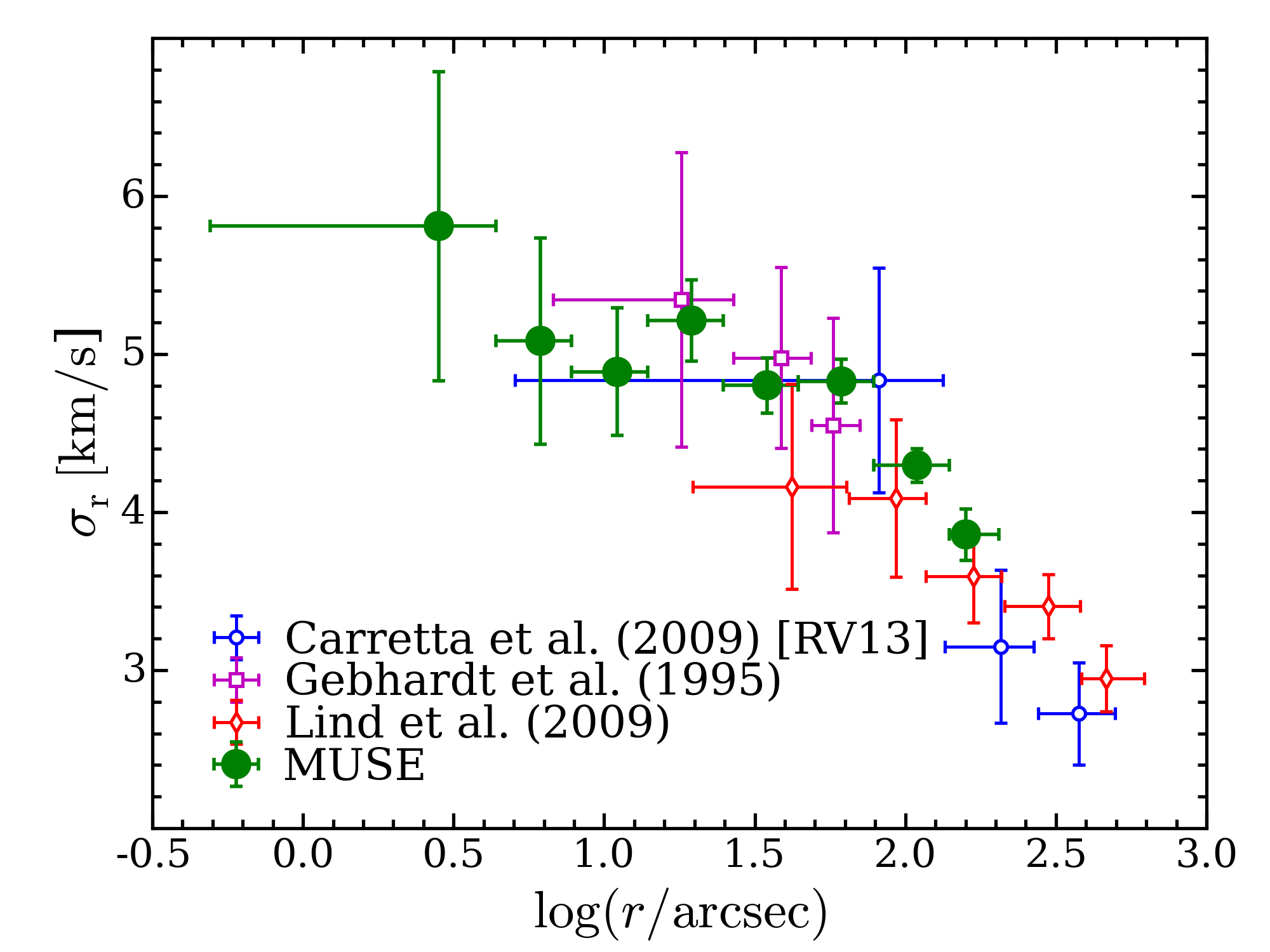}
  \caption{The velocity dispersion profile of \object{NGC~6397} as it was derived from the MUSE data (green filled circles) is compared to profiles obtained using different literature sources (open symbols; see legend for details).}
  \label{fig:dispersion_comparison}
 \end{figure}

 To visualise the variation of the velocity dispersion with distance from the cluster centre, we combined the velocities in radial bins, where each bin was constructed such that it contained at least 50 stars and covered a radial range of $d\log(r/\arcsec)\geq0.2$. In each bin, the velocity dispersion was determined using the maximum likelihood method introduced by \citet{1993ASPC...50..357P}, which also takes the uncertainties of the velocity measurements into account when deriving the intrinsic dispersion of the cluster. The same approach was used to obtain velocity dispersion profiles from the literature data introduced in Sect.~\ref{sec:literature}. The individual profiles are compared in Fig.~\ref{fig:dispersion_comparison}, which shows  excellent agreement between our profile and the literature studies in the regions where the data overlap. However, as a result of the much larger stellar sample obtained with MUSE, the uncertainties in our dispersion profile are significantly smaller than in the other profiles. The finding that we can reliably determine velocity dispersions as low as $3.5\ \mathrm{km\ s^{-1}}$ confirms that we have a very good handle on the error budget in the MUSE data and that the wavelength accuracy provided by the instrument is remarkably high given its complexity and rather low spectral resolution.
 
 The central bins of the MUSE velocity dispersion curve suggest a small central cusp, although a flat profile is still within the range of our measurement uncertainties. None of the literature data sets probe the kinematics so close to the centre, so it is not surprising that the comparison profiles do not reproduce this feature. However, very recently \citet{2015ApJ...803...29W} published a dispersion profile for \object{NGC~6397,} based on proper motions obtained with HST, which is consistent with being flat in its central part.
 
\begin{figure*}
  \includegraphics[width=17cm]{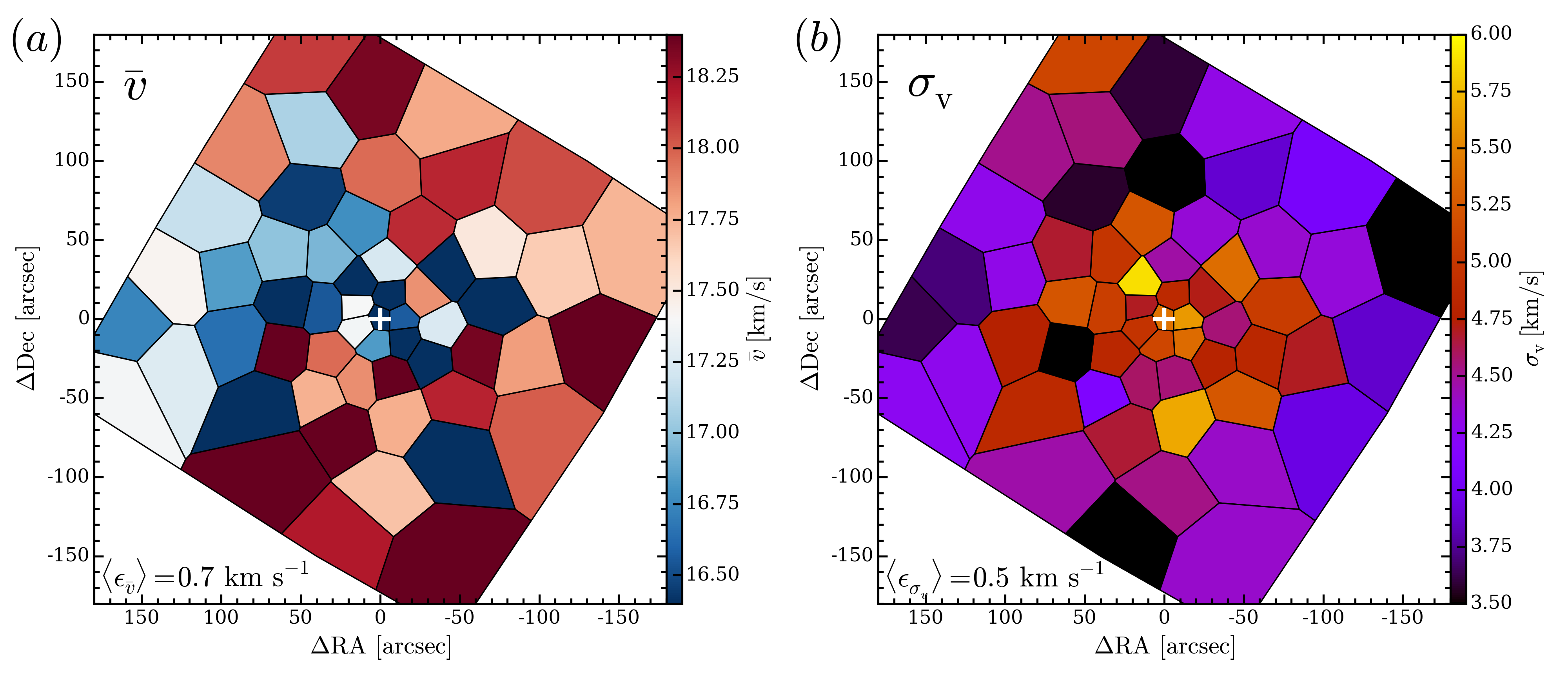}
  \caption{Two-dimensional kinematics of \object{NGC~6397} across the footprint covered by the MUSE observations, obtained by binning the data via Voronoi tesselation. The median velocity of the cluster stars is shown in panel (a), while panel (b) gives their velocity dispersion in each bin. In both panels, a white cross indicates the cluster centre and the average uncertainties per bin are provided in the lower left corner.}
  \label{fig:2d_kinematics}
 \end{figure*}

 We also investigated the velocity dispersion of the cluster in 2D. To this aim, we adapted an approach similar to that used by \citet{2015ApJ...803...29W} to derive a two-dimensional velocity dispersion map. Using the tool of \citet{2003MNRAS.342..345C}, we performed a Voronoi tesselation of our data. First, the data were binned on a grid with a cell size of $10\arcsec\times10\arcsec$. Each cell was then assigned a ``signal'' of $N$ and a ``noise'' of $\sqrt{N}$, where N was the number of data points in that cell. Finally, the Voronoi tesselation was performed on this grid  to achieve a final binning with approximately $N=100$ stars per bin. The result of this computation is shown in Fig.~\ref{fig:2d_kinematics}b. As for the 1D profile, a slight gradient from the centre of the cluster to the outskirts of the MUSE footprint can be identified. An advantage of the 2D profile is that it allows one to identify pecularities such as asymmetries in the velocity field or an offset between the kinematic and  photometric centre. However, we do not see evidence for either of these effects in our data.
 
 Encouraged by the accurate results we obtained for the overall dispersion curve, we checked whether more information can be extracted from the MUSE data, and in particular, whether we can detect any variation in the measured dispersion profile with stellar mass. As mentioned earlier, mass segregation has been observed in \object{NGC~6397}, which is a consequence of energy equipartition, i.e. the expectation that gravitational encounters between stars tend to equalise the kinetic energies of  participating stars. Hence, lighter stars should move faster on average.
 
\begin{figure*}
  \includegraphics[width=17cm]{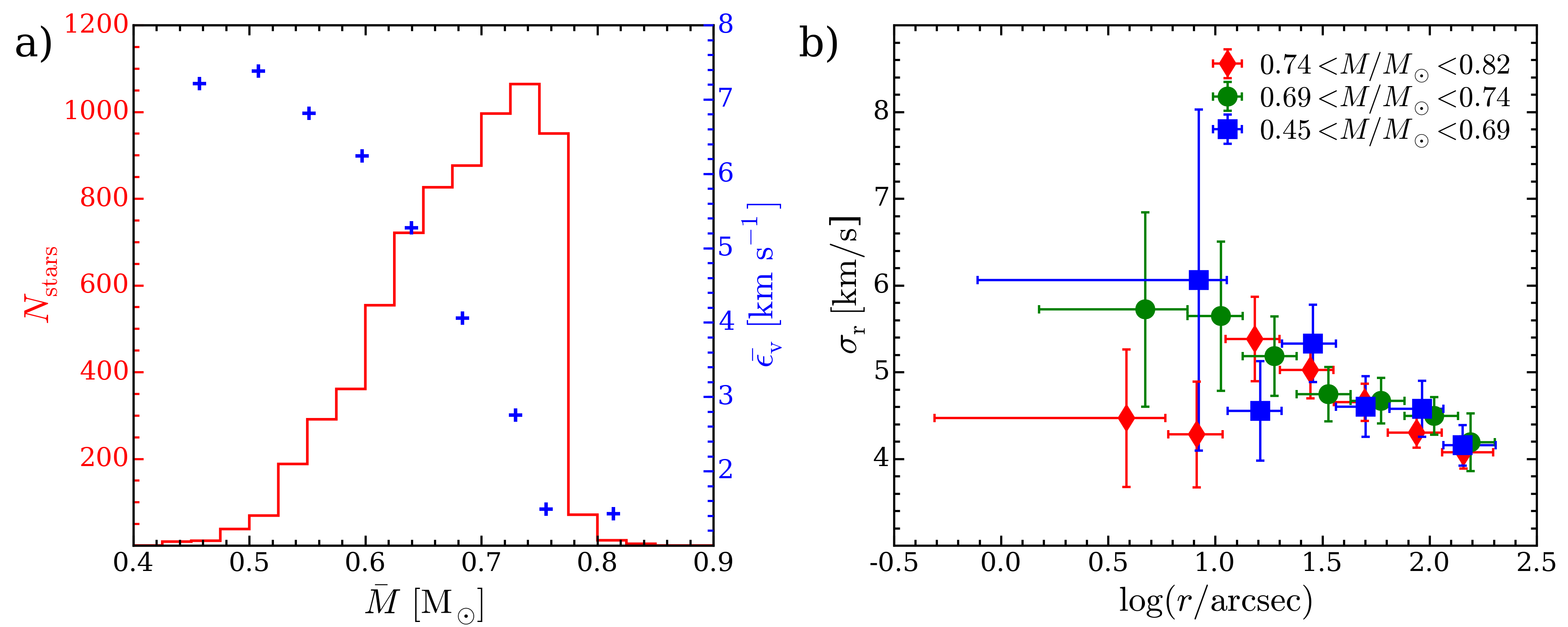}
  \caption{Analysis of the mass-dependent kinematics of \object{NGC~6397}. (a) The red histogram shows the distribution of stellar masses in the MUSE sample. The median measurement uncertainties of the stellar velocities as a function of mass are shown as blue crosses. (b) The velocity dispersion profile as a function of stellar mass. The mass range of each bin is shown in the legend. In contrast to (a), only stars with a measurement uncertainty $<5\ \mathrm{km\ s^{-1}}$ were included in this comparison.}
  \label{fig:dispersion_mass}
 \end{figure*}

 For this type of analysis, we require masses for the stars in our data set. In \citetalias{2015A&A...subm....H} we illustrated how initial guesses for the spectral fits were determined by comparing the photometry of the individual stars to a theoretical isochrone of the cluster, which was obtained from the database of \citet{2012MNRAS.427..127B}. The isochrone also allows us to assign a mass to each star from the MUSE sample. In doing this, we find that the masses of the stars range from $0.45M_\odot$ to $0.82M_\odot$ (cf. Fig.~\ref{fig:dispersion_mass}a). Mass variations by a factor of ${\sim} 1.8$ are large enough to search for signs of energy equipartition. However, a major obstacle for this investigation is that the mass of a star approximately scales with its luminosity, hence, on average stars with a higher mass are extracted with a higher S/N from the MUSE data. As a consequence, the mass of a star correlates strongly with the uncertainty of its radial velocity measurement. Fig.~\ref{fig:dispersion_mass}a shows that while the median uncertainty for a star with $0.8M_\odot$ is $1.5\,\mathrm{km\,s^{-1}}$, this value increases to $7.5\,\mathrm{km\,s^{-1}}$ for a star with $0.5M_\odot$. While we are confident that our uncertainties are accurate, we wanted to avoid any artificial signal caused by residuals in the determination of our error budget. Therefore, we restricted ourselves to stars which have a measurement uncertainty of $\epsilon_\mathrm{v_r}<5\ \mathrm{km\ s^{-1}}$ (about $2/3$ of our sample). While this selection does not remove the trend that more massive stars have more accurate velocity measurements, it does reduce the impact of the measurement uncertainties on the analysis.
 
 The velocity dispersion profiles obtained for the individual mass bins are shown in Fig.~\ref{fig:dispersion_mass}b. In the outskirts of the cluster, the profiles are more or less indistinguishable from one another. This finding agrees with the simulations carried out by \citet{2013MNRAS.435.3272T} in which energy equipartition is only partially developed. This result can be understood as a consequence of the instability that in a two-mass system the massive stars form a self-gravitating system if their mass fraction exceeds a critical value \citep{1969ApJ...158L.139S}.
 
 At the cluster centre, i.e. at radii $\lesssim10\arcsec$, the situation may be different, as we observe the highest mass bin to have a lower dispersion than the other mass bins with a difference of about $1\,\mathrm{km\,s^{-1}}$. This is comparable to the uncertainties of the individual datapoints, hence the significance of the observed difference is low. Regarding the results on partial energy equipartion we just mentioned, a spread between the mass bins would not be expected, in particular because the mass difference between the upper two bins is not large. On the other side, the observations carried out by \citet{2014MNRAS.442.3105M} showed that the strongest mass segregation is observed close to the cluster centre. Unfortunately, our data set does not include enough low-mass stars close to the cluster centre to study their kinematics at a sufficient level of detail to investigate this further, and we conclude that more data are needed to settle this issue.
 
 So far it was only possible to investigate energy equipartition with HST \citep[e.g.][]{2013MNRAS.435.3272T} because of the limited sizes and mass ranges covered by radial velocity samples. However, our results at larger distances to the cluster centre show that  this restriction can be overcome with MUSE and that it may become possible to investigate this matter at much lower observational costs. In this respect, it should be highlighted that here we are still working on commissioning data with short exposure times. With its high throughput, the light-collecting power of the VLT, and our capability to deblend stellar spectra, MUSE can dig significantly deeper into the stellar populations of a cluster than we demonstrated in this study, provided that longer exposure times are being used. Close to the centre, however, the observations are limited by crowding rather than by exposure time. Increasing the spatial resolution, for example via the adaptive optics module that is planned to support future MUSE observations, will be  key. 

 \subsection{Rotation}
 \label{sec:rotation}

 In order to check if we detected any significant rotation in \object{NGC~6397}, we used the same Voronoi bins as in Sect.~\ref{sec:dispersion} and determined the median velocity in each bin. The result of this calculation is shown in Fig.~\ref{fig:2d_kinematics}a. We detected no strong rotational signal in \object{NGC~6397}, but there is a trend towards lower velocities in the eastern region of the observed footprint compared to the western region with a difference of about $1\mathrm{km\ s^{-1}}$. This would hint of a weak rotational component around an axis that is approximately aligned in the north-south direction. In addition, we also observe that the median velocities seem to be lower near the cluster centre than in the outskirts of the observed footprint. However, just as the amplitude of the potential rotation curve, the strength of this feature is comparable to the expected residuals of the wavelength calibration (cf. Sect.~\ref{sec:uncertainties}). Therefore, the significance is hard to quantify. We checked the influence of the bin size by varying $N$ between $50$ and $200$. This did not change the overall appearance shown in Fig.~\ref{fig:2d_kinematics}a.
 
 Based on an analysis of the integrated light, \citet{1995AJ....110.1699G} found the core of \object{NGC~6397} to be weakly rotating with a projected amplitude ${\sim}2\mathrm{km\ s^{-1}}$. On the other hand, when looking at individual stars no significant rotational component was found. Still, weak rotations seem to be common in globular clusters. This is suggested by the study of \citet{2014ApJ...787L..26F}, who reported rotational signatures to be ubiquitous in the eleven (northern) clusters they analysed. The signals detected by \citeauthor{2014ApJ...787L..26F} show gradients of ${\sim}1\mathrm{km\ s^{-1}/arcmin}$, comparable to what is suggested by our analysis of \object{NGC~6397}. We conclude that if \object{NGC~6397} is rotating at all, the rotational component is very weak and that the velocity dispersion is clearly the dominant contribution to the second order velocity moment. Hence, we neglect rotations in the following discussion of Jeans modelling and refer to the second order velocity moment simply as velocity dispersion $\sigma_\mathrm{r}$.

 \subsection{Jeans modelling}
 \label{sec:jeans}
 
  \begin{figure*}
  \includegraphics[width=18cm]{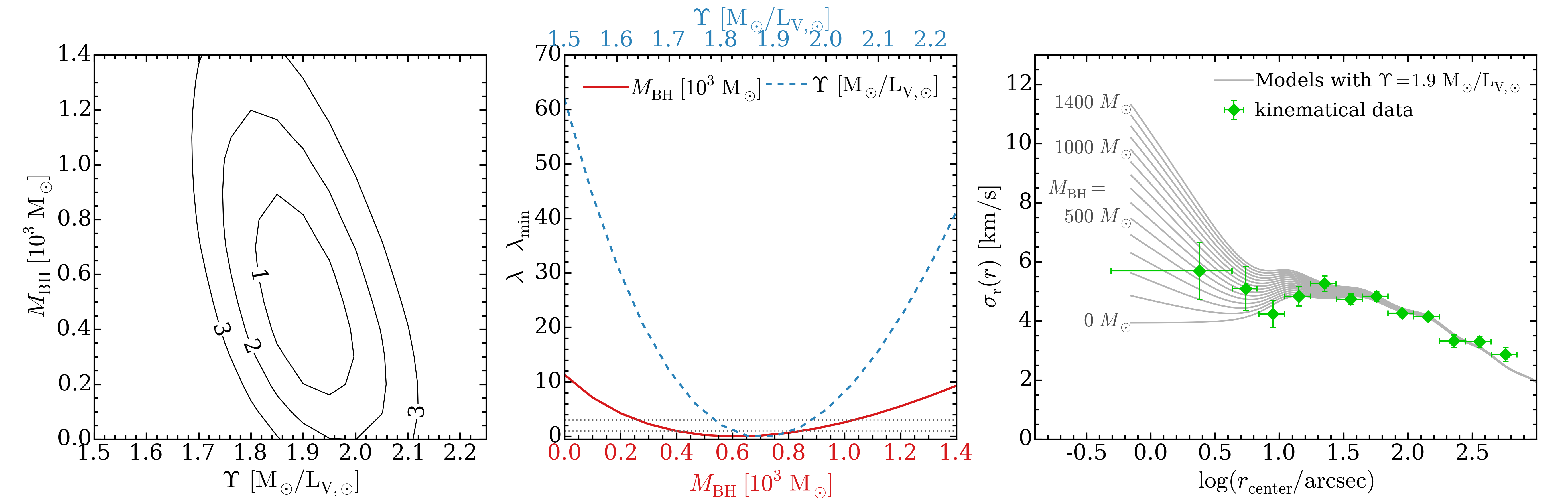}
  \caption{Comparison between the kinematical data and the predictions of spherical Jeans models with a constant mass-to-light ratio $\Upsilon$ and different masses $M_\mathrm{BH}$ for a central black hole. The left panel shows the distribution of likelihoods in the $\Upsilon$-$M_\mathrm{BH}$ plane. Solid lines give $1$, $2$, and $3\sigma$ confidence intervals. The marginalised distributions of both $\Upsilon$ and $M_\mathrm{BH}$ are shown in the central panel. The dotted lines indicate the $1$ and $3\sigma$ confidence intervals. The right panel compares the velocity dispersion predicted by models with a mass-to-light ratio of $\Upsilon=1.9\,\mathrm{M}_\odot/\mathrm{L}_\odot$ and different black hole masses to the kinematical data.}
  \label{fig:jeans_constantml}
 \end{figure*}
 
 Jeans modelling is a widely used method to infer cluster properties from photometric and kinematical data. In the spherical case, which is appropriate to \object{NGC~6397} since we did not find strong signs of rotation in our analysis of Sect.~\ref{sec:rotation}, the Jeans equation is written as \citep[e.g.][]{2008gady.book.....B}
 \begin{equation}
  \frac{d(\nu\overline{v_\mathrm{r}^2})}{dr} + \frac{2 \beta\,\nu\overline{v_\mathrm{r}^2}}{r} = -\nu \frac{d\Phi}{dr}\,.
 \end{equation} 
 It relates the gravitational potential $\Phi$ of the cluster to the second order central velocity moment $\overline{v_\mathrm{r}^2}$, weighted by the luminosity density $\nu$. The parameter $\beta = 1 - \overline{v_\mathrm{\theta}^2}/\overline{v_\mathrm{r}^2}$ measures the anisotropy of the velocity distribution inside the cluster.
 
 A common approach is to start from the surface brightness profile of the cluster and obtain its luminosity density via deprojection of the measured profile. Under the assumption of a given mass-to-light ratio (M/L), we can obtain an estimate of the gravitational potential of the cluster. The contribution of a central black hole or a population of dark stellar remnants may be added to this potential. The Jeans equation relates the gravitational potential to the second order moments of the velocity distribution of the cluster stars. In practice, the second order moments are predicted for a range of cluster parameters, such as M/L or the black hole mass, and the model that best matches the data is inferred by means of a $\chi^2$ or maximum-likelihood test. For our analysis of \object{NGC~6397} we used the JAM modelling code of \citet{2008MNRAS.390...71C}. This code requires the surface brightness profile of the cluster to be parametrised as a multi-Gaussian expansion, which is what we did in Sect.~\ref{sec:sbprofile}.
  
 To compare the model predictions of the velocity dispersion to our measurements, we used the same maximum-likelihood approach as in \citet{2002AJ....124.3270G} and \citet{2014A&A...566A..58K}. This approach has the advantage that it does not require any binning of the data. Instead, for each star $i$ one measures the probability of drawing its radial velocity, $v_i$, from the predicted velocity distribution at radius $r_i$. The likelihood $\mathcal{L}$ of a model given the data is then calculated as the product of the probabilities of all stars. It can be shown that for large data sets, the quantity $\lambda=-2\ln\mathcal{L}$ follows a $\chi^2$ distribution. The implication of this for our data analysis is twofold.
 \begin{enumerate}
  \item We consider a model as a statistically valid representation of our data if the deviation between its $\lambda$ value and the expectation value $\langle\lambda\rangle$ is within a given confidence interval that is determined from a $\chi^2$ distribution with $N$ degrees of freedom. In this case, $N$ is the number of stars in the data.
 \item We use likelihood ratio tests to determine confidence intervals for the model parameters. If the most likely model yields $\lambda=\lambda_{\rm min}$, the quantity $\Delta\lambda=\lambda-\lambda_{\rm min}$ follows a $\chi^2$ distribution with $m$ degrees of freedom. In this case, $m$ is the number of free parameters in the model. Unless otherwise noted, the confidence intervals we provide are $1\sigma$ intervals.
  \end{enumerate}
 The data used in this comparison is the union of all data shown in Fig.~\ref{fig:dispersion_comparison}, after a combination of the results obtained for the individual stars and an exclusion of RV variables which resulted in a final sample of $N=7027$ stars.

 The first suite of models that we calculated had a constant mass-to-light ratio $\Upsilon$, hence possible effects of mass segregation were neglected. We varied $\Upsilon$  in the range $1.6\leq\mathrm{M}_\odot/\mathrm{L}_\odot\leq2.4$ with a step size of $0.1\,\mathrm{M}_\odot/\mathrm{L}_\odot$. A central black hole with a mass varying between $0$ and $1\,500\ \mathrm{M}_\odot$ in steps of $100\ \mathrm{M}_\odot$ was included in the models. The left panel of Fig.~\ref{fig:jeans_constantml} shows the 2D likelihood distribution of the model parameters given our data. The most likely model has $\Upsilon=1.9\,\mathrm{M}_\odot/\mathrm{L}_\odot$ and contains a black hole with $500\,\mathrm{M}_\odot$. The likelihood of this model deviates by ${\sim}1\sigma$ from the expected value for $N=7027$, hence it provides a statistically valid description of our data.
 
 For further analyses, we determined marginal distributions for both parameters, which are depicted in the central panel of Fig.~\ref{fig:jeans_constantml}. We obtained a value for the (dynamical) M/L of $\Upsilon = 1.9\pm0.1\mathrm{M}_\odot/\mathrm{L}_\odot$, which is in good agreement with values determined by \citet[][$2.1\pm0.1\mathrm{M}_\odot/\mathrm{L}_\odot$]{1991A&A...250..113M} and \citet[][$2.4\pm0.5\,\mathrm{M}_\odot/\mathrm{L}_\odot$]{2012ApJ...761...51H}. Using the absolute cluster magnitude obtained in Sect.~\ref{sec:sbprofile}, we measured a total cluster mass of $M_\mathrm{tot}=(7.0\pm0.8)\cdot10^4\ M_\mathrm{\odot}$, which is about $30\%$ smaller than the estimates of \citet{1991A&A...250..113M} and \citet{2012ApJ...761...51H}. The unknown aspect in our mass estimate is the accuracy of the integrated magnitude of the cluster, as already discussed in Sect.~\ref{sec:rvvariables}. An offset of $30\%$ in mass corresponds to a magnitude offset ${\sim}0.25$, which seems not unreasonably high in view of the large scatter found for the integrated magnitude in the literature.
 
 In the right panel of Fig.~\ref{fig:jeans_constantml}, we compare the velocity dispersion profiles predicted by models with a constant M/L of $\Upsilon=1.9\mathrm{M}_\odot/\mathrm{L}_\odot$ and various black hole masses to our measurements. It is remarkable how well the models can reproduce the velocity dispersion profile that we measured. One can also see that the model without a black hole underestimates the measured central dispersion. Without a black hole, a slight decrease of the velocity dispersion within $10\arcsec$ is expected. Hence our models with a constant M/L require a black hole to match the data. We obtain a best-fit mass of $M_\mathrm{BH}=600\pm200\,\mathrm{M}_\odot$.
 
 In Fig. 7, we observe a decrease in velocity dispersion towards the cluster centre predicted by the model without any black hole in our highest stellar mass bin. In this respect, it would be very interesting to obtain model predictions for the kinematics as a function of stellar mass. We cannot perform such modelling with our approach, but it is possible for example with N-body models. However, such modelling is beyond the scope of this paper.
 
 Profiles that show a decreasing velocity dispersion towards the centre are a common result of isotropic models, i.e. models with $\beta\equiv0$. As anisotropies can provide another explanation for a rise in the central dispersion profile, we tested what degree of central anisotropy would be required to achieve a similar effect compared to a black hole with $600\,\mathrm{M_\odot}$. This was carried out via comparison of our data to a second grid of Jeans models, where instead of including a black hole the $\beta$-value of the central Gaussian components of the surface brightness profile was altered. We found that a strong radial anisotropy of $\beta=0.5$ would be required. This would strongly contradict the results of the  proper motion study by \citet{2015ApJ...803...29W}, who did not find any evidence for central anisotropies. This result is in agreement with the expectation that central anisotropies are suppressed by the relaxation of the cluster. In addition, the kinematics in the outskirts of the cluster are also consistent with $\beta\equiv0$ \citep{2012ApJ...761...51H}. Therefore, we do not expect that setting $\beta\equiv0$ oversimplifies our models.
 
 Still the question is whether a black hole is the only way to bring the data and the models into agreement. For example, \citet{2002AJ....124.3270G} found in their study of \object{M~15} that the necessity for a black hole was diminished when a varying M/L was assumed, which in this case was obtained from Fokker-Planck simulations. These simulations predicted a rise in the central M/L, mainly caused by the segregation of stellar remnants near the centre. Mass segregation has also been observed in \object{NGC~6397}, so we wanted to investigate the influence of a varying M/L on our data-model comparison. While the prediction of a dedicated M/L profile via Fokker-Planck or N-body simulations is beyond the scope of this paper, we instead followed the approach of \citet{2010ApJ...710.1063V} and added an additional gravitational component to account for a central increase in the density of dark stellar remnants. To this aim, we modelled its density via the simple profile introduced by \citet{1911MNRAS..71..460P}  to describe the distribution of stars in globular clusters. This profile is well constrained in extent and provides a simple representation for a compact mass characterised by a single scale. Denoting the total mass $M_\mathrm{d}$ and the length scale $a_\mathrm{d}$, its functional form is written as
 \begin{equation}
  \label{eq:plummer}
  \rho(r) = \frac{3\cdot M_\mathrm{d}\cdot a_\mathrm{d}^2}{4\pi\cdot(r^2 + a_\mathrm{d}^2)^{5/2}}\,.
 \end{equation}

 In view of the good agreement between our models and our data in the outskirts of the cluster, we kept the M/L of the luminous stellar component constant at $\Upsilon=1.9\ \mathrm{M}_\odot/\mathrm{L}_\odot$ and created models for a range in masses ($0 < M_\mathrm{d}/M_\odot < 3\,000$) and length scales ($0\farcs3 < a_\mathrm{d} < 30\arcsec$) but without a central black hole. The \citeauthor{1911MNRAS..71..460P} profile was parametrised using a multi-Gaussian expansion and added as an additional density component into the Jeans modelling code. The surface brightness profile input into the code was left unchanged because the dark component does not contribute to it.

\begin{figure}
  \includegraphics[width=\hsize]{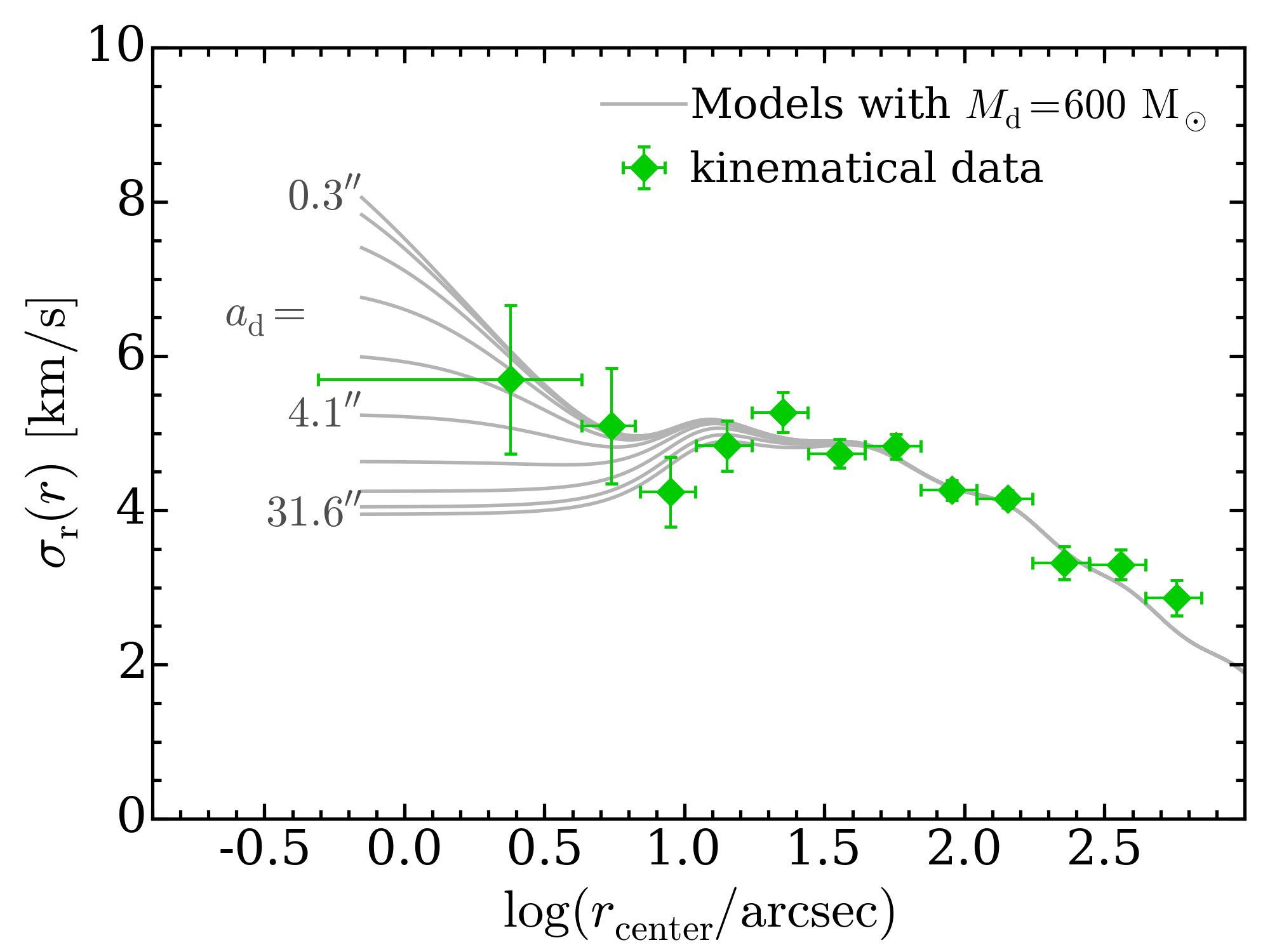}
  \caption{Comparison between Jeans models with a variable mass-to-light ratio and the kinematical data. The grey curves correspond to models that have a central dark stellar component with a mass of $600\,\mathrm{M_\odot}$ and characteristic scales between $0\farcs3$ and $30\arcsec$ as indicated in the plot.}
  \label{fig:jeans_varyingml}
 \end{figure}

 The general observation when carrying out this analysis is that a central accumulation of remnants can also explain our measurements. However, the models that provide the best match to our data contain dark components that closely resemble the black hole, i.e. have similar masses and are compact. This is illustrated in Fig.~\ref{fig:jeans_varyingml}, where we compare our data to a suite of models with a constant mass of $M_\mathrm{d}=600\,\mathrm{M}_\odot$ and different length scales $a_\mathrm{d}$. These models only improve the agreement with the data for values $a_\mathrm{d}\lesssim5\arcsec$ compared to the case with a constant M/L and no black hole.
 
 Statistically, the most likely model with varying M/L, which has $M_\mathrm{d}=600\,\mathrm{M}_\odot$ and $a_{\rm d}=2.45\arcsec$, represents the data as well as the most likely model containing a central black hole. For this reason, the data at hand do not allow us to exclude one of the two options: an intermediate-mass black hole or a compact overdensity of remnants. In the next section, we discuss whether either of the two alternatives can be excluded by other means.
 
 \section{Discussion and conclusions}
 \label{sec:conclusions}
 
 The comparison of our kinematical data with the predictions from the isotropic Jeans models shows that the central velocity dispersion is higher than what could be explained by the bright stars alone. An intermediate-mass black hole with a mass of $600\pm200\,\mathrm{M_\odot}$ would be an intriguing explanation for this. If its presence is confirmed, it would be the first detection of an intermediate-mass black hole based on the analysis of individual stars instead of integrated light. Using the central velocity dispersion of $\sigma=5\,\mathrm{km\,s^{-1}}$ and the mass of $M=7\cdot10^4\,\mathrm{M_\odot}$ of the cluster, we can check if the black hole would fall onto the scaling relations known from galaxies (always taking into account that this involves the extrapolation of a power law over several orders of magnitude). The relations derived by \citet{2003ApJ...589L..21M} for the mass and by \citet{2009ApJ...698..198G} for the velocity dispersion both yield black hole masses of ${\sim}100\,\mathrm{M_\odot}$.  However, as argued by \citet{2013MNRAS.434L..41K}, instead of using the current properties of the clusters, one may rather use their initial properties when applying the scaling relations. The reason is that the relaxation times of stellar clusters are much shorter than those of galaxies, hence while galactic bulges have not experienced much evolution during their lifetimes, globular clusters have. Interestingly, when we apply the predictions of \citet[their Figs.~1 and 2]{2013MNRAS.434L..41K} to \object~{NGC6397}, we obtain an expected black hole mass of ${\sim}1\,000\,M_\mathrm{\odot}$.
 
 We find that a compact accumulation of dark stellar remnants provides an alternative explanation for the kinematical data we have obtained. The main contribution to this component would probably be in the form of stellar black holes and neutron stars because their masses are significantly higher than those of the bright stars (in contrast to most white dwarfs). Estimations on the total mass of neutron stars in \object{NGC~6397} based on simulations have been obtained by \citet{1995ApJS..100..347D} or \citet{2009MNRAS.395.1173G}, yielding $>1\,400\,\mathrm{M_\odot}$ and $3\,140\,\mathrm{M_\odot}$, respectively. The total mass of the dark stellar component that provides the best fit to our data is only $600\,\mathrm{M_\odot}$. From this point of view, the budget that is available in the cluster would be sufficient to create such a central dark component. Observationally, however, the abundance of dark remnants is not well constrained. The sample sizes obtained in gamma- or X-ray observations of globular clusters are relatively small, with ${\sim}10$--$100$ candidate objects \citep[e.g.][]{2010A&A...524A..75A}. In \object{NGC~6397}, only one neutron star has been identified by its X-ray emission \citep{2001ApJ...563L..53G}. In addition, a few companions of millisecond pulsars have been detected \citep[e.g.][also \citetalias{2015A&A...subm....H}]{2001ApJ...561L..93F}.
 
 Another aspect concerns the question of how long such a dense central component is stable before it evaporates via gravitational encounters between its members or tidal forces. For an old ($>10\,\mathrm{Gyr}$) stellar system, the following constraint has been derived for the half-mass radius $r_\mathrm{h}$ of a central component that has not evaporated yet \citep[][eq. 7.143]{2008gady.book.....B}:
 \begin{equation}
  \label{eq:evaporation}
  r_\mathrm{h} \geq 0.01\,\mathrm{pc}\cdot\left(\ln(0.1\cdot M/m)\right)^{2/3}\left(\frac{m}{M_\odot}\right)^{2/3}\left(\frac{10^8M_\odot}{M}\right)^{1/3}\,,
 \end{equation}
 where $M$ is the total mass of the component and $m$ the average mass of its constituents. Using $M\equiv M_\mathrm{d}=600\,\mathrm{M_\odot}$ and assuming $m=1\,\mathrm{M_\odot}$, eq.~\ref{eq:evaporation} yields $r_\mathrm{h} \geq 1.4\,\mathrm{pc}$. For a \citeauthor{1911MNRAS..71..460P} profile, the integration of eq.~\ref{eq:plummer} shows that $r_\mathrm{h}\approx1.3\,a_\mathrm{d}$. Therefore, the value for $r_\mathrm{h}$ corresponds to $a_\mathrm{d,evap} \geq 90\arcsec$ at a distance of $2.5\,\mathrm{kpc}$, more than a factor of $10$ larger than our upper limit on the scale radius, $a_\mathrm{d}=5\arcsec=0.06\,\mathrm{pc}$. We conclude that any central component formed during the early evolution of the cluster would have been evaporated by now unless the average mass of its constituents was $m\ll1\,\mathrm{M_\odot}$, in which case its collapse via collisions may become an issue \citep[e.g.][]{1995ApJ...447L..91M}.
 
 However, in a globular cluster a central component would rather form via mass segregation than as a remnant of the formation of the cluster. The evaporation timescale is directly linked to the relaxation time $t_\mathrm{h}$ which scales with $r_\mathrm{h}^{3/2}$ \citep[][eq. 7.108]{2008gady.book.....B}. Setting $r_\mathrm{h}=1.3\,a_\mathrm{d}=0.09\,\mathrm{pc}$, we obtain an evaporation timescale of ${\sim}160\,\mathrm{Myr}$. This is longer than the estimated core relaxation time of the cluster, $t_\mathrm{c}\approx0.1\,\mathrm{Myr}$ \citep{1996AJ....112.1487H}, which governs the mass segregation. Hence it is possible that the central component is quasi-stable because it is constantly refilled with massive objects from the outer cluster regions.
 
 By design, our models with an additional \citeauthor{1911MNRAS..71..460P} components in their centres would all cause a rise in the central M/L. However, the results from previous studies cast doubts on whether that is the case, even if the remnants accumulate in the centre. \citet{2014MNRAS.442.3105M} investigated mass segregation among the luminous stars in \object{NGC~6397} and found that their mass-to-light ratio \emph{decreased} towards the centre; while the average mass of a star increased from $0.55\,\mathrm{M}_\odot$ at $r=30\arcsec$ to $0.8\,\mathrm{M}_\odot$ near the core radius, its average magnitude decreased by ${\sim}3\,\mathrm{mag}$ in the same range. The combined effect of luminous stars and remnants was investigated by \citet{1991A&A...250..113M}, who modelled the dynamics of the cluster using multi-mass King-Mitchie models. While the authors found that indeed massive neutron stars have the highest concentration, their contribution was not strong enough to make up for mass segregation among luminous stars. Hence M/L dropped from its global value of $2.0\,\mathrm{M}_\odot/\mathrm{L}_\odot$ to a value of $1.7\,\mathrm{M}_\odot/\mathrm{L}_\odot$ around the centre.
 
 The surface brightness profile of \object{NGC~6397} may provide further evidence about which of the two aforementioned scenarios, a central black hole or an accumulation of stellar remnants, is more likely. As we mentioned earlier, the cluster may have undergone core collapse, even if our analysis of the surface brightness profile is also consistent with a flattening within $2\arcsec$ around the centre. The simulations of \citet{2011ApJ...743...52N} have shown that it is unlikely to find intermediate-mass black holes in core collapse clusters. In addition, \object{NGC~6397} does not show the large ratio between core radius and half-mass radius that \citet{2007MNRAS.374..857T} found for simulated globular clusters containing an IMBH. 
 
 A puzzling observation is the dependence of the measured velocity dispersion on the mass of the stars near the centre of the cluster, although the significance is low. If this is an intrinsic effect, it would be difficult to explain with energy equipartition in view of the results by \citet{2013MNRAS.435.3272T}. In particular because these authors found that an intermediate-mass black hole would suppress the equipartition of the cluster stars even further. However, more data will be needed to investigate this further. In this respect, \object{NGC~6397} is actually not the optimal cluster to observe with MUSE. Its low velocity dispersion poses severe constraints on the accuracy of the measured velocities. At the same time, its low metallicity results in a lack of strong absorption lines, so that the spectra must be extracted with a higher S/N than in more metal-rich clusters to achieve the same accuracy.
 
 Still our analysis has shown the potential of crowded field 3D spectroscopy with MUSE to investigate the dynamics of Galactic globular clusters. We are currently carrying out a larger survey of several clusters. This survey focusses on massive clusters that have higher velocity dispersions than \object{NGC~6397, and therefore} we expect to learn even more from the analysis of the MUSE spectra. Also, a dedicated search for binary stars in the course of this survey will be important in addressing several questions that remain in the current paper. For example, luminous stars orbiting around massive stellar remants should be relatively easy to detect, hence it would be possible to find an accumulation of dark stellar remnants near the cluster centre.
 
 \section*{Acknowledgements}
 SK and PMW received funding through BMBF Verbundforschung (project MUSE-AO, grant 05A14BAC and 05A14MGA). SK further acknowledges support from DFG via SFB 963/1 ``Astrophysical flow instabilities and turbulence'' (project A12).
 
 \bibliographystyle{aa}
 \bibliography{aa_2015_27065}
  
\end{document}